\documentclass[12pt,epsfig]{article}
\usepackage{graphicx,amssymb}
\usepackage[figuresright]{rotating}

\newcommand{\noi}{\noindent}
\newcommand{\eq}{\begin{equation}}
\newcommand{\en}{\end{equation}}
\newcommand{\eqa}{\begin{eqnarray}}
\newcommand{\ena}{\end{eqnarray}}
\def\Journal#1#2#3#4{{#1}{\bf #2} (#4) #3}

\def\NPB{Nucl. Phys.~{\bf B}}
\def\NPPS{Nucl. Phys. Proc.~Suppl.~}

\def\PLB{Phys.~Lett.~{\bf B}}

\def\PRD{Phys.~Rev. ~{\bf D}}

\def\ZPC{Z.~Phys.~{\bf C}}

\begin{document}

\title{On Dirac sheet  configurations\\ of  $SU(2)$ lattice  fields}
\author{
E.-M. Ilgenfritz$^a$, B. V. Martemyanov$^b$, M. M\"uller-Preussker$^a$ \\
 and A. I. Veselov$^b$ \\ $~$ \\
{\small
$^a$ Humboldt-Universit\"at zu Berlin, Institut f\"ur Physik,
D-10115 Berlin, Germany} \\
{\small
$^b$ Institute for Theoretical and Experimental Physics,
Moscow 117259, Russia}
}

\maketitle

\vspace*{-12cm}
\begin{flushright}
{\small ITEP-LAT-2003-03\\ HU-EP-03/72 }
\end{flushright}
\vspace*{10cm}

\begin{abstract}
Finite temperature Euclidean $SU(2)$ lattice gauge fields
close to the deconfinement phase transition are subjected
to cooling. We find relatively stable or absolutely stable
configurations with an action below the one-instanton 
action $S_{inst}=2\pi^2$ both in the deconfinement and the confinement phases.
In this paper we attempt an interpretation of these lowest action
configurations. Their action is purely magnetic and amounts to
$S/S_{inst} \approx N_t/N_s$, where $N_t$ ($N_s$) is the
timelike (spacelike) lattice size, while the topological charge
vanishes. In the confined phase part of the corresponding lattice
configurations turns out to be absolutely stable with respect to
the cooling process in which case Abelian projection reveals a
homogeneous, purely Abelian magnetic field closed over the "boundary"
in one of the spatial directions.
Referring to the dyonic structure established for the confinement phase
near $T_c$ and based on the observation made for this phase that such
events below the instanton action $S_{inst}$ emerge from dyon-antidyon annihilation,
the question of stability (metastability) is discussed for both phases.
The hypothetically different dyonic structure of the deconfinement phase,
inaccessible by cooling, could explain the metastability.
\end{abstract}

\section{Introduction}
\label{sect:introduction}
In the confined phase, below but sufficiently close to $T_c$,
applying cooling to Monte Carlo generated $SU(2)$ lattice gauge fields
has shown a dyonic structure of metastable action
plateaus~\cite{IMPV,IMMPV1,IMMPV2,IMMPSV}.
The dyons themselves correspond, in a good approximation, to the
constituents of Kraan-van Baal-Lee-Lu (KvBLL) caloron 
solutions~\cite{KvB,big_paper,LL}.
The same cooling applied to lattice configurations in the deconfined
phase acts entirely differently. There is no remnant dyonic structure.
Instead of this, there are metastable events on some lowest action
plateau (actually significantly below the one-instanton action 
$S_{inst}=2\pi^2$). 
These configurations have vanishing topological density, and the 
action is purely magnetic. Such configurations have been already 
observed and discussed many years ago in papers by Laursen and 
Schierholz~\cite{LS}, and Veselov and Polikarpov~\cite{VP}.

In the present paper, from the perspective gained with the dyonic
structure at high enough temperature, we try to give a new interpretation
of the lowest action configurations seen in Refs.~\cite{LS,VP}. 
For this goal we study them both in confined and deconfined phases. 
In the confinement phase we are in the fortunate position that we 
can observe the parent configurations which sometimes evolve into 
the configurations considered here. In all observed cases this parent 
configuration was a dyon-antidyon pair. Contrary to this, in the 
deconfined phase the cooling technique has been unable to exhibit 
potential parent configurations in the form of action plateaus, 
i.e. approximate solutions of the lattice equations of
motion (metastable plateaus of action). 
We have critically examined the 't Hooft-Polyakov monopole structure
of the cooled configurations emerging from the deconfinement phase
that was suggested in Ref.~\cite{LS}. We do confirm to see a minor
fraction of magnetic configurations which resemble 
't Hooft-Polyakov monopoles at a first view. However, even for these 
rare events we find only a weak correlation between the
localization of magnetic action and the positions of monopoles defined 
either in terms of the Polyakov line or of the magnetic charge in the 
Weyl gauge. 
 
On the contrary, the similarity of the action dependence for these 
configurations on the spatial size of the lattice suggestively points 
toward their common nature. Most likely, in both cases we would expect 
quantized magnetic fluxes.
The returned flux (unavoidable in the maximal Abelian projection for periodic
boundary conditions and manifesting itself as a Dirac string) can be 
visualized as Dirac sheet (swept out by a Dirac string moving in 
Euclidean time). Hence, we adopt the name Dirac Sheets (DS) for this 
class of observed cooled lattice configurations. DS configurations are 
known as exact solutions of the lattice field equations in $U(1)$ 
LGT~\cite{Mitr}.

The paper is organized as follows.

In Section \ref{sect:formulae} we will provide all necessary lattice 
definitions, in particular the observables considered in order to identify 
KvB and DS solutions.

In Section \ref{sect:ds} we report on the statistics and the properties of DS
events observed both in the confined and the deconfined phases.

Section \ref{sect:conclusions} contains our conclusions.

\section{Production and characterization of DS solutions}
\label{sect:formulae}
Throughout this paper $SU(2)$ gauge theory in four-dimensional Euclidean
space is considered on an asymmetric lattice with periodic boundary 
conditions in all four directions. The respective ensembles of 
configurations have been created by heat bath Monte Carlo using 
the standard Wilson plaquette action
\eqa \label{eq:action}
S = \sum_{\vec{x},t} s(\vec{x},t) &=& \sum_{\vec{x},t} \sum_{ \mu < \nu}
                    s(\vec{x},t;\mu,\nu), \\
s(\vec{x},t;\mu,\nu) &=& \beta~(1 - \frac{1}{2}\mathrm{tr}~U_{x,\mu\nu}),\qquad
               U_{x,\mu\nu} = U_{x,\mu} U_{x+\hat{\mu},\nu}
               U^{\dagger}_{x+\hat{\nu},\mu} U^{\dagger}_{x,\nu} \nonumber
\ena
with inverse coupling $\beta=4/g_0^2$. For simplicity the lattice spacing
is set equal to $a=1$. The lattice size was $N_s^3 \times N_t$ with the
spatial extension $N_s = 8,10,12,16,20$ and with the inverse physical
temperature $T^{-1} \equiv N_t = 4$. For $N_t=4$ the model is known to
undergo the deconfinement phase transition at the critical coupling
$\beta_c \simeq 2.299$ \cite{Bielefeld}. Throughout this paper we will use
two ensembles with $\beta =2.2 < \beta_c$ and  $\beta =2.4 > \beta_c$.

The equilibrium field configurations in both ensembles have been cooled
by iterative minimization of the Wilson action $S$. In one or another form,
cooling is used in order to smooth out short-range fluctuations, while
(initially) leaving untouched some large-scale properties of the configurations.
The cooling method applied here is the standard relaxation method described
long time ago in \cite{ILMPSS}
and was used for investigation of instantons \cite{ILMPSS,PV}.

This method, if applied without any further limitation, easily finds
approximate solutions of the lattice field equations as shoulders
(plateaus) of the action as a function of cooling steps (relaxation
history). Here we shall concentrate on smoothed fields at the very last
stages of cooling, using a stopping criterium which selects the plateaus
in the interval of action $ S  \le 0.6 ~S_{\mathrm{inst}}. $.

The emerging gauge field configurations were analyzed according to
the spatial distributions of the following observables:

\begin{itemize}

\item {\it action density} computed from the local plaquette values:
  \eq \label{eq:actdensity}
s(\vec{x},t)= \sum_{ \mu < \nu} s(\vec{x},t;\mu,\nu) \, ,
  \en
  (see eq.(\ref{eq:action}));

\item {\it topological density} computed with the standard
twisted plaquette discretization:
  \eq \label{eq:topdensity}
  q_t(\vec{x}) = -  \frac{1}{2^4 \cdot 32 \pi^2}
                \sum_{\mu,\nu,\rho,\sigma=\pm 1}^{\pm 4}
           \epsilon_{\mu\nu\rho\sigma}
           \mathrm{tr} \left[ U_{x,\mu\nu} U_{x,\rho\sigma} \right]
           \,;
  \en

\item {\it Polyakov loop} defined as:
  \eq \label{eq:latpolloop}
  L(\vec{x}) = \frac{1}{2} \mathrm{tr} P
  \en
  with
  \eq
  P(\vec{x}) = \prod_{t=1}^{N_t} U_{\vec{x},t,4} \, ,
  \en
  where the $U_{\vec{x},t,4}$ represent the links in time direction;

\item {\it non-stationarity } defined as:
  \eq
  \delta_t =\sum_{\vec{x},t} |s(\vec{x},t+1)-s(\vec{x},t)|/S~;
  \en

\item {\it violation of equations of motion} $\Delta$
\eq
\Delta = \frac{1}{4 N_s^3 N_t} \sum_{x,\mu} \frac{1}{2}
         \mathrm{tr} \left[ (U_{x,\mu}-\bar{U}_{x,\mu})
                            (U_{x,\mu}-\bar{U}_{x,\mu})^{\dagger}
                                            \right]\,,
\en
where
\eq \nonumber
\bar{U}_{x,\mu}= c \sum_{\nu \ne \mu} \left[
U_{x,\nu}  U_{x+\hat{\nu},\mu} U^{\dagger}_{x+\hat{\mu},\nu} +
U^{\dagger}_{x-\hat{\nu},\nu} U_{x-\hat{\nu},\mu} U_{x+\hat{\mu}-\hat{\nu},\nu}
                                    \right]
\en
is the local link $x,\mu$ being the solution of the lattice equation of
motion, with all degrees of freedom coupled to it held fixed.
The factor $c$ is just a normalization of the staple sum such that
$\bar{U}_{x,\mu} \in SU(2)$ \footnote{The replacement
$U_{x,\mu} \rightarrow \bar{U}_{x,\mu}$ is exactly
the local cooling step as applied throughout this paper.}.

\item {\it Abelian magnetic fluxes and monopole charges} defined within
the Weyl (or generalized temporal) gauge, $\partial_0 A_0=0$,
and the maximally Abelian gauge (MAG).
The first one is obtained from the Polyakov gauge (PolG, achieved by 
diagonalizing $P(x)$ in each lattice site, followed by Abelian gauge
transformations which render the (then diagonal) temporal links 
$U_{\vec{x},\tau;4}$ independent on time $\tau$. The latter is found by 
maximizing the gauge functional $A$ 
  \eq \label{eq:mag}
  A[g] = \frac{1}{2}~\sum_{x,\mu}
       \mathrm{tr} ( U^{g}_{x,\mu} \tau_3 U^{g \dagger}_{x,\mu} \tau_3 )\,,
  \en
  under gauge transformations $U_{x,\mu} \rightarrow U^g_{x,\mu}
  = g(x) U_{x,\mu} g^{\dagger}(x + \hat{\mu})$. 
  The abelianicity is the maximum value of this quantity divided by
  the number of links.

  In both cases, Abelian link angles $\theta_{x,\mu}$ are then defined
  by Abelian projection onto
  the diagonal $U(1)$ part of the link variables $U_{x,\mu} \in SU(2)$.
  According to the DeGrand-Toussaint prescription \cite{dGT}
  a gauge invariant magnetic flux $\bar{\Theta}_p$ through an oriented
  plaquette $p \equiv (x,\mu\nu)$
  is defined by splitting the plaquette
  $\Theta_p = \theta_{x,\mu} +\theta_{x+\hat{\mu},\nu}
              -\theta_{x+\hat{\nu},\mu}-\theta_{x,\nu}$
  into
  $\Theta_p =\bar{\Theta}_p + 2\pi n_p, ~~n_p=0,\pm 1,\pm 2$ such that
  $\bar{\Theta}_p \in (-\pi, +\pi]$.
  The magnetic charge of an
  elementary 3-cube $c$ is then
  $m_c = \frac{1}{2\pi} \sum_{p \in \partial c} \bar{\Theta}_p\,.$

\end{itemize}

\noi
For the cooling procedure of equilibrium gauge field configurations
we have kept the standard periodic boundary conditions on the $4D$ torus.

Finally, cooling was stopped at some ($n$-th) cooling iteration step when
the following criteria for the action $S_n$ were simultaneously fulfilled:
\begin{itemize}
\item $S_n < 0.6~S_{\mathrm{inst}}$,
\item $S_n-2~S_{n-1}+S_{n-2} < 0$.
\end{itemize}

\noi
The last condition means that the relaxation just passed a point of inflection.
As we have empirically observed, the point of inflection always coincides,
within an accuracy of plus/minus one global cooling step, with a minimal
violation of the equations of motion $\Delta$.
This can be understood as follows.
If violation of equations of motion $\Delta=0$, i.e.
equations of motion are fulfilled we are in the local minimum of the action
where its variation is zero. If violation of equations of motion $\Delta$
has the minimum the variation of the action has also the minimum and second
variation of the action is zero what means that the action goes through
the point of inflection.

\section{Properties of Dirac Sheets in the confined and the deconfined phases}
\label{sect:ds}
\subsection{Cooling}

\begin{figure}[!htb]
\begin{center}
\includegraphics[width=.7\textwidth]{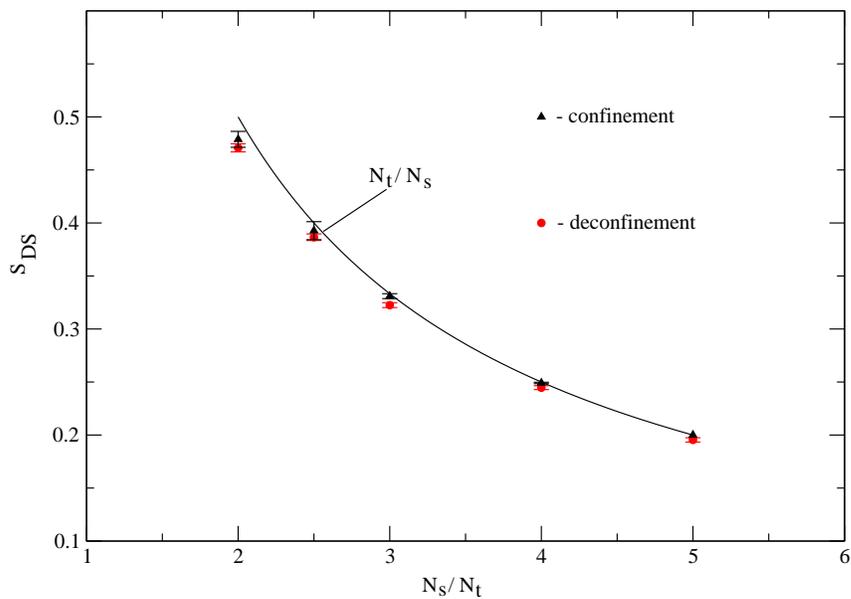}
\caption{Action of DS events in the confined (triangles) and in the deconfined 
(circles) phases for lattices with $N_t=4$ and $N_s=8,10,12,16,20$ as function 
of $N_s/N_t$.}
\label{fig:ds_act_nsnt03}
\end{center}
\end{figure}

\begin{figure}[!htb]
\begin{center}
\includegraphics[width=.7\textwidth]{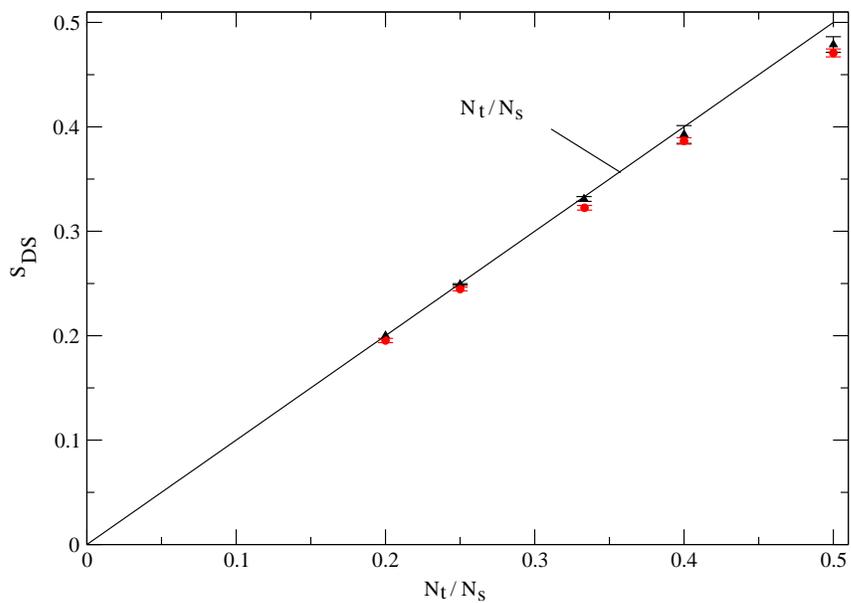}
\caption{Same as in Fig. \ref{fig:ds_act_nsnt03} as function of $N_t/N_s$.}
\label{fig:ds_act_ntns03}
\end{center}
\end{figure}

We have investigated DS events on lattices $N_s^3 \times N_t$
with $N_t=4$ and $N_s=8,10,12,16,20$.
The statistics and the mean actions of various types of configurations 
selected by the cooling process are presented in Table 1 and 
Figures \ref{fig:ds_act_nsnt03} and \ref{fig:ds_act_ntns03} 
where the dependence of ${\bar S}/S_{inst}$ on $ N_s/N_t$ and $ N_t/N_s$
is shown to have the tendency ${S_{DS}}/S_{inst}\rightarrow N_t/N_s$.
Their properties are summarized in Table 2 and 
illustrated by the remaining Figures.
We have found  DS becoming very stable at action values
$\simeq S_{\mathrm{inst}}\cdot  N_t/N_s$.
The (color-)electric contribution to the total action is very small compared
with the magnetic contribution.
Moreover, they are perfectly static with the values of nonstaticities
$\delta_t$ shown in Table 2.
Employing MAG we have convinced ourselves that they are almost Abelian
(see the abelianicity $A$ in Table 2), and only a small fraction
contains MAG monopoles. 

In the confinement phase it happens quite rarely that they appear {\it directly}
in the result of the cooling process. In all cases observed they appear
after dyon-antidyon pairs have been observed at
$S \approx S_{\mathrm{inst}}$ which annihilate in the final stage of relaxation.
The Abelian monopole content of DS in the deconfinement phase, if it is 
obtained by Abelian projection in the Weyl gauge, amounts to
monopole-antimonopole pairs being present in $60\div 90 \%$ of those cooled
configurations.

We originally had found this type of solution for fixed holonomy boundary 
conditions (f.h.b.c.) \cite{IMPV,IMMPV1,IMMPV2,IMMPSV}.
These DS for f.h.b.c. were seen to be oriented very exactly in plane and 
to have non-zero action for plaquettes in one of the space-space coordinate 
planes $(x,y)$, $(x,z)$ or $(y,z)$.
They had the same action values $S/S_{\mathrm{inst}}$ as we have found 
lateron for the case of periodic boundary conditions.
We had also found configurations which contained two DS orthogonal 
to each other at action values twice as large than for one DS. We shall
not further comment on such events in this paper.

\begin{table}[ht]
\caption{The statistics of DS events on the lattices
$N_s^3 \times N_t$ with $N_s=8,10,12,16,20$
for $\beta=2.20 $ (confined phase, upper row) and $\beta=2.40 $
(deconfined phase, lower row). $\bar{S}$ denotes the average
action of the observed events, $\delta S$ the variance.}
\vspace{5mm}
\begin{center}
\begin{tabular}{llccc}
\hline
$\beta $ & $ N_s/N_t$ & $ N_{event} $ & ${\bar S}/S_{inst}$ & $\delta{\bar S}/S_{inst}$ \\
\hline
$ 2.2 $ & $ 2.0 $ & $ 33 $ & $ 0.479 $ & $ 0.007 $  \\
$ 2.4 $ & $ 2.0 $ & $ 54 $ & $ 0.471 $ & $ 0.004 $  \\
\hline
$ 2.2 $ & $ 2.5 $ & $ 18 $ & $ 0.392 $ & $ 0.009 $  \\
$ 2.4 $ & $ 2.5 $ & $ 62 $ & $ 0.387 $ & $ 0.003 $  \\
\hline
$ 2.2 $ & $ 3.0 $ & $ 12 $ & $ 0.331 $ & $ 0.002 $  \\
$ 2.4 $ & $ 3.0 $ & $ 73 $ & $ 0.323 $ & $ 0.002 $  \\
\hline
$ 2.2 $ & $ 4.0 $ & $ 11 $ & $ 0.249 $ & $ 0.0007 $ \\
$ 2.4 $ & $ 4.0 $ & $ 66 $ & $ 0.245 $ & $ 0.002  $ \\
\hline
$ 2.2 $ & $ 5.0 $ & $  7 $ & $ 0.200 $ & $ 0.000  $ \\
$ 2.4 $ & $ 5.0 $ & $ 28 $ & $ 0.195 $ & $ 0.002  $ \\
\hline

\end{tabular}
\end{center}
\end{table}
\noi
\begin{table}[ht]
\caption{The properties of DS events on the lattices
$N_s^3 \times N_t$ with $N_t=4$, $N_s=8,10,12,16,20$.
For each $N_s$ the events are presented in three rows:
unstable (upper row) and absolutely stable (middle row) 
DS in the confined phase for $\beta=2.20$ and DS in 
deconfined phase for $\beta=2.40$ (lower row). }
\vspace{5mm}
\begin{center}
\begin{tabular}{lcccccc}
\hline
$ N_s/N_t$ & frequency of & fraction of DS & fraction of DS 
                                           & A & $ \Delta$ & $\delta_t $ \\
           & DS events    & with monopoles & with monopoles 
                                           &   &           &             \\
           &              & in Weyl gauge  & in MAG         
                                           &   &           &             \\
\hline
2.0&3.8\%&21.1\%&5.3\%&98.8\%&0.223E-05&0.372E-03\\
2.0&3.2\%&-&-&100\%&0.944E-09&0.253E-06\\
2.0&4.4\%&59.1\%&6.8\%&98.0\%&0.355E-05&0.889E-03\\
\hline
2.5&3.0\%&41.7\%&8.3\%&98.9\%&0.122E-05&0.426E-03\\
2.5&4.8\%&-&-&100\%&0.167E-15&0.245E-12\\
2.5&11.3\%&79.4\%&2.9\%&98.8\%&0.107E-05&0.512E-03\\
\hline
3.0&2.0\%&50.0\%&-&99.5\%&0.378E-06&0.214E-03\\
3.0&5.7\%&-&-&100\%&0.150E-15&0.389E-12\\
3.0&14.3\%&88.4\%&-&99.1\%&0.464E-06&0.397E-03\\
\hline
4.0&0.3\%&-&-&99.9\%&0.385E-10&0.103E-07\\
4.0&5.0\%&-&-&100\%&0.760E-15&0.108E-10\\
4.0&16.0\%&72.9\%&14.6\%&99.5\%&0.665E-07&0.149E-03\\
\hline
5.0&-&-&-&-&-&-\\
5.0&2.8\%&-&-&100\%&0.636E-16&0.245E-11\\
5.0&18.0\%&77.8\%&11.1\%&99.7\%&0.244E-07&0.951E-04\\
\hline
\end{tabular}
\end{center}
\end{table}
\noi
\subsection{Can DS configurations be viewed as 't Hooft-Polyakov \\
monopoles in the deconfinement phase ?}
 
In contrast to previous parlance (these configurations have been called 
''monopole'' ($M$) configurations in Refs.~\cite{LS,IMMPV2}) we have called 
them here Dirac Sheet (DS) configurations from the beginning. Now we want
to provide some more facts supporting this interpretation. In the beginning
let us recall several features which originally suggested the interpretation
as 't Hooft-Polyakov monopoles.

There should be  
\begin{itemize}
\item a minimum of the dynamically generated ''Higgs field'' 
      $\mathrm{tr} (A_0/T)^2$
\item a maximum of the (almost purely) magnetic action density
      $\mathrm{tr} B^2$
\item a pair of pointlike Abelian magnetic charges, one carrying 
      action and the other being spurious.
\end{itemize}.

We have studied these signatures for our configurations cooled down below the
one-instanton level starting from the deconfinement phase, {\it i.e.} from 
equilibrium Monte Carlo configurations on a $16^3\times 4$ lattice at 
$\beta=2.4$. All these configurations have one maximum of the magnetic
action and a relatively shallow maximum of the Polyakov line. A subset
of these configurations, considered in the Weyl or maximally Abelian gauge, 
respectively, has a pair of corresponding static magnetic charges. 
The relative frequencies actually to find the indicated Abelian monopole 
structure is given in Table 2.

\begin{figure}[!htb]
\begin{center}
\includegraphics[width=.4\textwidth,height=0.3\textheight]{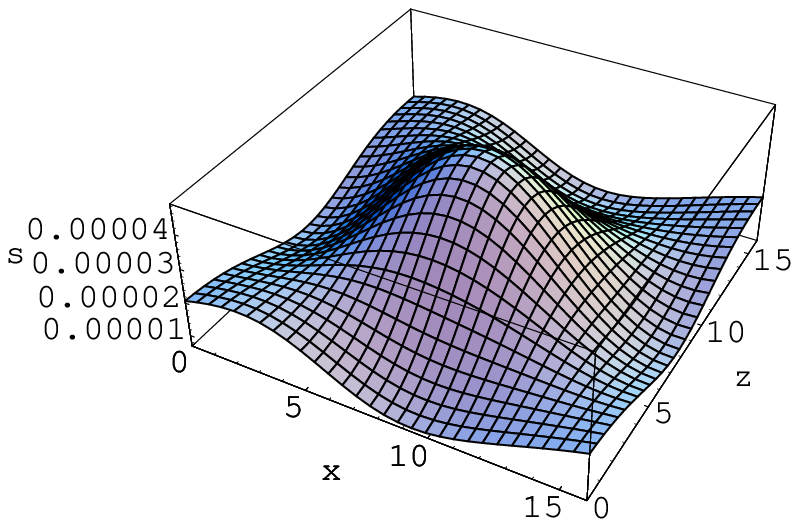}%
\hspace{.3cm}
\includegraphics[width=.4\textwidth,height=0.3\textheight]{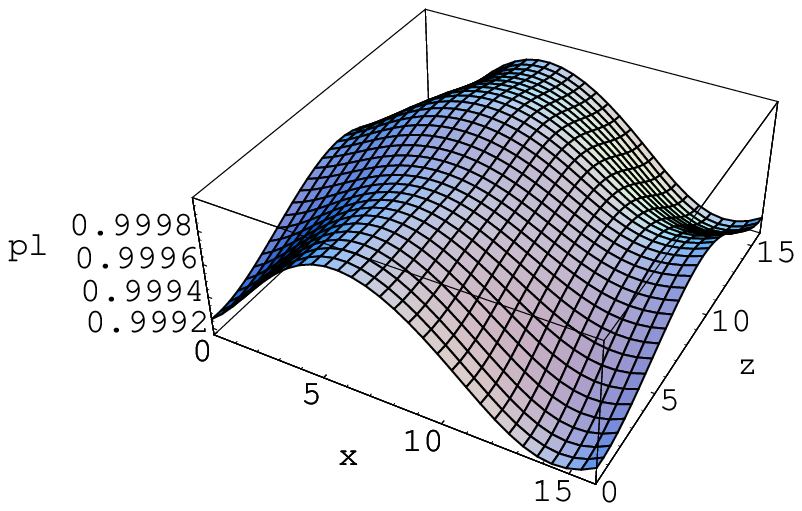}
\end{center}
\begin{center}
\caption{Action density (left) and Polyakov line (right) of an DS event 
in the deconfined phase shown as function over the $xy$ plane cutting
through the maximum of action density.}
\label{fig:act_pol}
\end{center}
\end{figure}

\begin{figure}[!htb]
\begin{center}
\includegraphics[width=.5\textwidth,height=0.3\textheight]{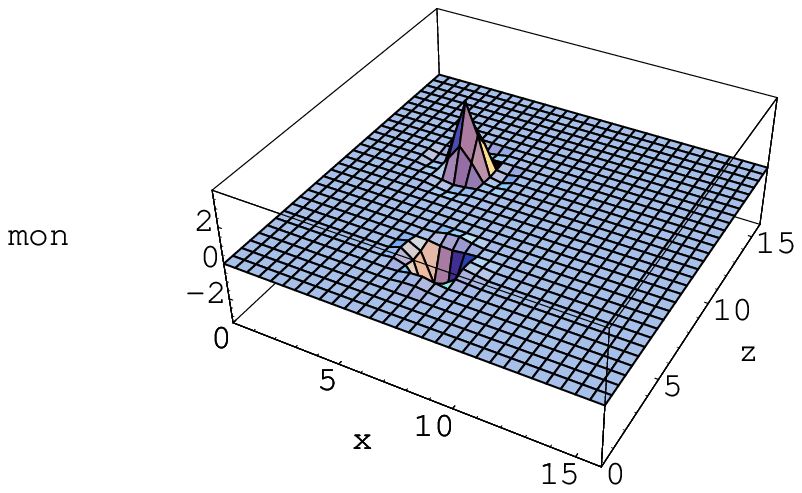}
\caption{Abelian monopole charge density in Weyl gauge (summed over $z$) of 
the DS event presented on Fig.\ref{fig:act_pol}.
For static monopoles the monopole charge density is equal to $\pm N_t=\pm 4$.
The smearing is due to the interpolation in MATHEMATICA graphics.}
\label{fig:P-monopoles}
\end{center}
\end{figure}

One of the cooled configurations with a Weyl gauge monopole pair is 
portrayed in Figure \ref{fig:act_pol} where we show the profiles of 
action density and Polyakov line over the $(x,z)$ plane which cuts 
the configuration and contains the maximum of action density. 
In Figure \ref{fig:P-monopoles} the positions of 
the static monopole pair are projected onto that $(x,z)$ plane.
This event indeed shows a structure as reported in Ref. \cite{LS}.

For a more quantitative assessment, we have compared the positions of
the maxima of three-dimensional action density and the Polyakov loop. 
We found the average distance equal to $7 \pm 2$ lattice spacings,
whereas for the maximum of the action density and the position of the nearest 
Weyl gauge magnetic monopole the distances are somewhat less, $5.2 \pm 1.6$ 
lattice spacings.  These distances are certainly smaller than the maximal 
three-dimensional distance $8\sqrt{3} \approx 13$ but are difficult 
to reconcile with a  static extended particle (monopole) interpretation.

\begin{figure}[!htb]
\begin{center}
\includegraphics[width=.6\textwidth]{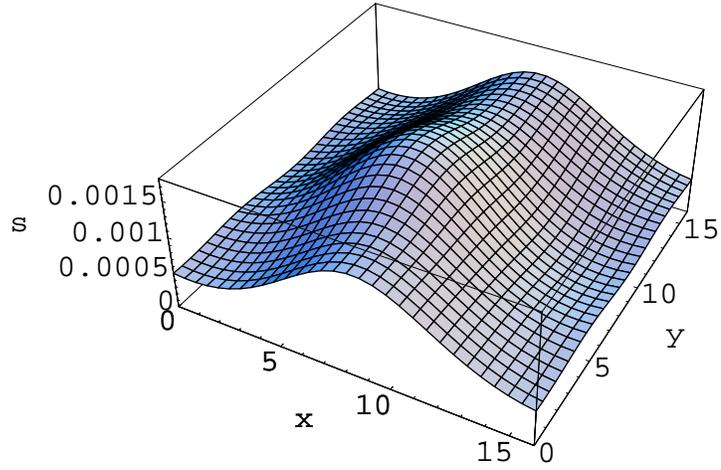}
\caption{Action density (summed over $t$ and $z$) of an DS event in 
the deconfined phase shown as function over the $xy$ plane.}
\label{fig:s_dsdec}
\end{center}
\end{figure}

\begin{figure}[!htb]
\begin{center}
\includegraphics[width=.6\textwidth]{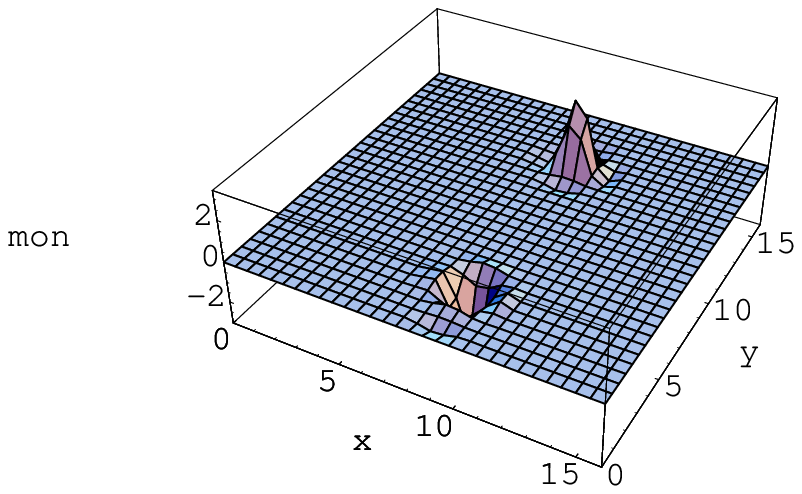}
\caption{Abelian monopole charge density in MAG (summed over $z$) of 
the DS event presented on Fig.\ref{fig:s_dsdec}.
For static monopoles the monopole charge density is equal to $\pm N_t=\pm 4$.
The smearing is due to the interpolation in MATHEMATICA graphics.}
\label{fig:mon_dsdec}
\end{center}
\end{figure}

Monopoles obtained in the MAG are present only in approximately
$< 15 \%$ of all DS events (see also Table 2).
One of the rare cases when a DS event in the deconfinement phase shows
an Abelian monopole-antimonopole pair in MAG is presented in
Figs.\ref{fig:s_dsdec} and \ref{fig:mon_dsdec}.

\subsection{More on the Dirac sheet interpretation}

At this point it might be useful to focus on the striking volume 
dependence of the action (common to both phases) and on the surprising 
stability of the cooled DS configurations (which distinguishes the 
confinement phase) in order to understand similarities and differences 
between the ways how these configurations originate from the respective
vacuum of the two phases.

The dependence of ${\bar S}/S_{inst}$ on $ N_s/N_t$ and $ N_t/N_s$
is shown on Figures \ref{fig:ds_act_nsnt03} and \ref{fig:ds_act_ntns03}.
In the confined phase for $N_s=20$ all 7 DS events were absolutely stable
with ${S_{DS}}/S_{inst}= N_t/N_s$.
The number  $N_t/N_s$ can be understood as follows. Let there be a quantized
homogeneous Abelian magnetic flux in some spatial ($x$, $y$ or $z$) direction.
The Abelian magnetic field is equal to $B^3_x=4\pi/N_s^2$. Its action is equal to
\cite{Mitr}
\eq
S_{DS}=\frac{4}{g^2_0}(1-\cos(\frac{B^3_x}{2})) N_s^3 N_t\approx \frac{1}{2g^2_0} {(B^3_x)}^2 N_s^3 N_t
=\frac{8\pi^2}{g^2_0}\frac{N_t}{N_s}=S_{inst}\frac{N_t}{N_s} \; .
\en
All 7 DS events in the confined phase for $N_s=20$ (when put into MAG) show
such an Abelian magnetic field.
There arises the question: why for other (smaller) $N_s$ in the confined phase
the fluxes are not always homogeneous and absolutely stable, and why the fluxes are
unstable for all $N_s$ in the deconfined phase.
The probable answer is that magnetic flux (during the process of $D\bar D$
annihilation) is not always closed over the "boundary" in some periodic
spatial direction.
If the size of dyons in the $D\bar D$ pair is small compared
to the spatial size of the lattice, the annihilation is almost pointlike and the
magnetic flux has a good chance to be closed.

The size distribution of dyons depends on holonomy.
In the confined phase the measure of
holonomy $L=\cos (2\pi\omega )$ ($\omega$ being the holonomy parameter) is
distributed in the neighbourhood of zero. For lower plateaus of action the
distribution approaches more and more the semicircle law (Haar measure).
Then $\omega$ is distributed over the range $0 \le \omega \le 1/2$
and the size of the lighter dyon as known from the KvB solution
($N_t/4\pi\omega$ for $0 \le \omega \le 1/4$ and
$N_t/4\pi(1/2-\omega)$ for $1/4 \le \omega \le 1/2$ measured in lattice spacings)
varies from $N_t/\pi$ to the maximal value possible on the finite lattice.
In the deconfined phase, the holonomy becomes closer and closer to the trivial one
($L\approx \pm 1$) and the dyons (in the dyon-antidyon pair) are strongly
delocalized.
As mentioned in the Introduction, during cooling in the deconfined phase the
dyon-antidyon pair itself does not become visible on a well-established plateau.

\begin{figure}[!htb]
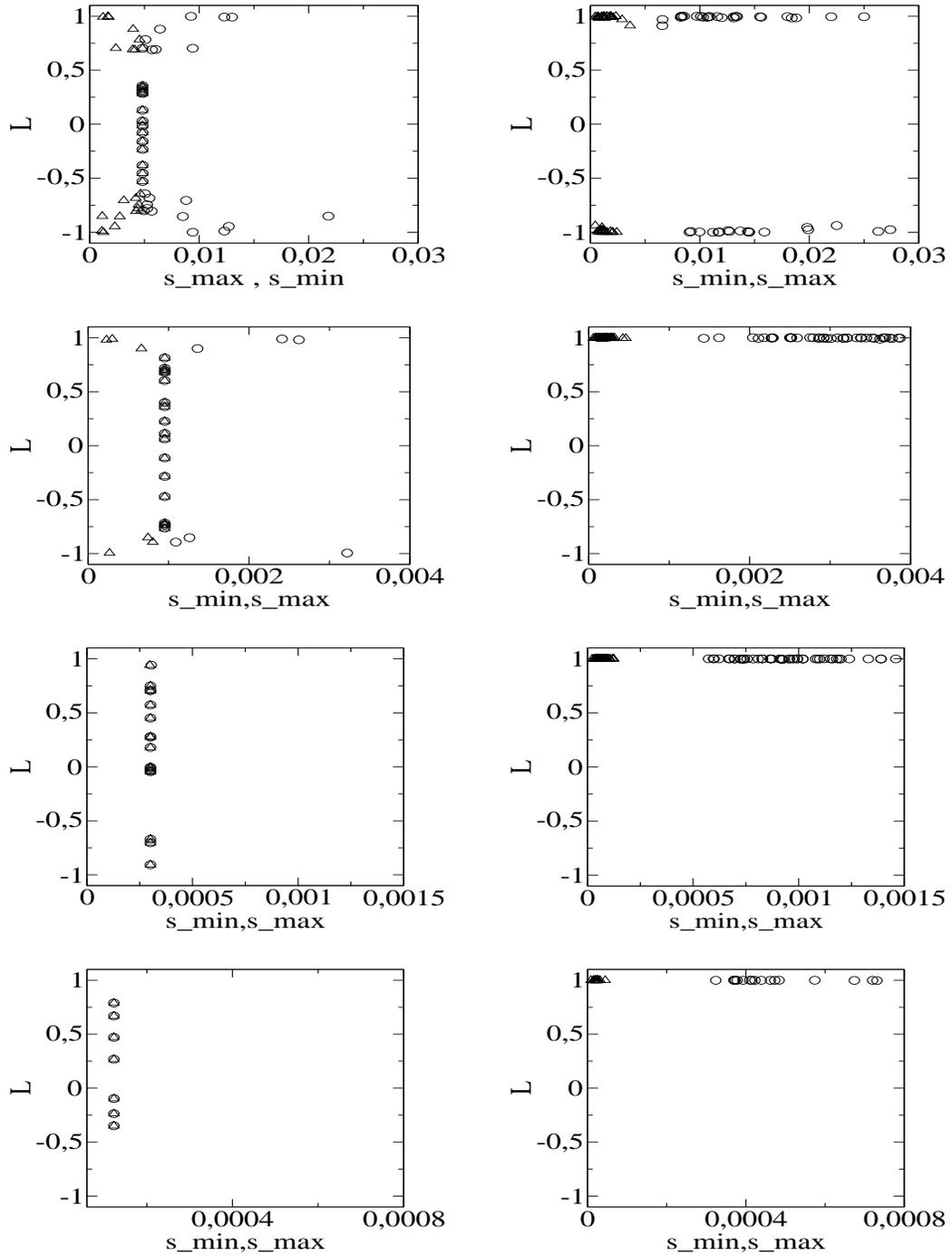

\begin{center}
\includegraphics[width=.4\textwidth,height=0.19\textheight]{dsconf8.eps}%
\hspace{.8cm}
\includegraphics[width=.4\textwidth,height=0.19\textheight]{dsdec8.eps}


\vspace{.5cm}
\includegraphics[width=.4\textwidth,height=0.19\textheight]{dsconf12.eps}%
\hspace{.8cm}
\includegraphics[width=.4\textwidth,height=0.19\textheight]{dsdec12.eps}

\vspace{.5cm}
\includegraphics[width=.4\textwidth,height=0.19\textheight]{dsconf16.eps}%
\hspace{.8cm}
\includegraphics[width=.4\textwidth,height=0.19\textheight]{dsdec16.eps}

\vspace{.5cm}
\includegraphics[width=.4\textwidth,height=0.19\textheight]{dsconf20.eps}%
\hspace{.8cm}
\includegraphics[width=.4\textwidth,height=0.19\textheight]{dsdec20.eps}

\caption {Correlation between the Polyakov loop $L$ and the stability of 
DS configurations, illustrated by scatter plots showing, for each DS event, 
two points: ($s_{min}$, $L$) (as triangles) and ($s_{max}$, $L$) (as circles)
with $s_{min}$ ($s_{max}$) being the minimal (maximal) action density of the 
DS configuration. The left coloumn presents DS events in the confined phase 
for lattices with $N_t=4$ and $N_s=8,12,16,20$ (from up to down),
the right coloumn presents DS events in the deconfined phase for the same 
lattice sizes.}
\label{fig:L-s_min-s_max}  
\end{center}
\end{figure}

The correlation between holonomy and stability of DS events is shown
in Figure \ref{fig:L-s_min-s_max} for both phases. 
It can be understood if the configurations are
really emerging from the annihilation of a dyon-antidyon pair.
The Figure presents scatter plots where each DS event is represented by two points:
($s_{min}$, holonomy) and ($s_{max}$, holonomy) with $s_{min}$ and $s_{max}$ being
the action density at sites where it is minimal and maximal, respectively).
Provided that the holonomy remains far enough from trivial, we obtain DS events 
from the confined phase which consist of homogeneous Abelian magnetic fluxes. 
The homogeneity is expressed by $s_{min}=s_{max}$ and
corresponds to the successful annihilation of more or less pointlike
dyon-antidyon pairs. However, for values of holonomy close to trivial holonomy
DS events in both confined and deconfined phase occur as unstable magnetic
fluxes which are not closed as the result of annihilation of less localized
(and ''massless'', i. e. low-action) dyons. The unstable DS in confined and
deconfined phases have similar characteristics as can be seen from the first
and third rows (shown separately for each $N_s$) in Table 2.

So, unstable DS events in confined and deconfined phases are similar.
There is no absolute gap between unstable and absolutely stable DS events in 
the confined phase.  This can be an argument in favor of their common nature.
The stable DS events found in the confined phase are purely Abelian magnetic
fluxes.

\section{Conclusions}
\label{sect:conclusions}
We have generated $SU(2)$ lattice gauge fields at non-zero temperature,
both in the confined and the deconfined phases. We have cooled them in order to
analyze the structure of the lowest action plateau (which in fact is below the
one-instanton action).  We have found certain structures ("Dirac sheets") that
resemble homogeneous Abelian magnetic fluxes.
The action dependence on the spatial lattice size $N_s$ favors such an
interpretation. 
For the deconfinement phase, where an 't Hooft-Polyakov monopole 
interpretation has been advocated, the loose correlation between
different possible definitions of how to localize the monopole as
an extended heavy particle makes this picture less convincing.
Instead, we have looked for another interpretation in terms of how
the original dyonic structure (which is different in the two phases) 
becomes destroyed by the cooling process.
In the infinite volume limit $N_s\rightarrow \infty$,
these DS structures disappear. Therefore, we interprete them as artefacts of 
the finite lattice volume.

\section*{Acknowledgements}
We gratefully acknowledge the kind hospitality of the Instituut-Lorentz of the
Universiteit Leiden extended to three of us (E.-M.I., B.V.M. and M.M.-P.) and
helpful discussions with Pierre van Baal and Falk Bruckmann when the revised 
version of this paper became completed.
This work was partly supported by RFBR grants 04-02-16079, 03-02-16941 and
02-02-17308, by the INTAS grant 00-00111, the CRDF award RP1-2364-MO-02,
DFG grant 436 RUS 113/739/0 and RFBR-DFG grant 03-02-04016,
by Federal Program of the Russian Ministry of Industry, Science and Technology 
No 40.052.1.1.1112. One of us (B.V.M.) gratefully appreciates the support of 
Humboldt-University Berlin where this work was initiated.
\newpage

%
\newpage


\vspace{1cm}

\vspace{1cm}

\vspace{1cm}

\end{document}